\documentclass[twocolumn,showpacs,floatfix,prl]{revtex4}
\usepackage{float}
\usepackage{graphicx}
\usepackage{amsfonts}

\begin{document}

\title{Odd-Parity Topological Superconductors: Theory and Application to Cu$%
_x$Bi$_2$Se$_3$}
\author{Liang Fu and Erez Berg}
\affiliation{Department of Physics, Harvard University, Cambridge, MA 02138}

\begin{abstract}
Topological superconductors have been theoretically predicted as a new class
of time-reversal-invariant superconductors which are fully gapped in the
bulk but have protected gapless surface Andreev bound states. In this work,
we provide a simple criterion that directly identifies this topological
phase in \textit{odd-parity} superconductors. We next propose a two-orbital $%
U-V$ pairing model for the newly discovered superconductor Cu$_x$Bi$_2$Se$_3$%
. Due to its peculiar three-dimensional Dirac band structure, we find that
an inter-orbital triplet pairing with odd-parity is favored in a significant
part of the phase diagram, and therefore gives rise to a topological
superconductor phase. Finally we propose sharp experimental tests of such a
pairing symmetry.
\end{abstract}

\maketitle


The search of topological phases of matter with time-reversal symmetry has
been an active field in condensed matter physics\cite{trs}. In the last few
years, a new phase of matter called topological insulators\cite{fkm, balents}
has been predicted\cite{fukane} and soon experimentally observed in a number
of materials\cite{bisbbise, shenbite}.
More recently, a new class of time-reversal-invariant (TRI) superconductors
has been predicted by a topological classification of Bogoliubov-de Gennes
(BdG) Hamiltonians\cite{ucsb, kitaev}. As a close cousin of topological
insulators, the so-called ``topological superconductor" is fully gapped in
the bulk but has gapless surface Andreev bound states hosting Bogoliubov
quasi-particles\cite{ucsb, stanford, roy}. Now the challenge is to
theoretically propose candidate materials for this new phase.

In this work, we first provide a simple criterion which can be directly used
to establish the topological superconductor phase in \textit{centrosymmetric}
materials with \textit{odd-parity} pairing symmetry. This criterion applies
to superconductors with spin-orbit coupling. 
We next study
the possibility of odd-parity pairing in the newly discovered superconductor
Cu$_x$Bi$_2$Se$_3$\cite{cubise}, which has a 3D Dirac band structure due to
strong SOC. We propose a phenomenological model for Cu$_x$Bi$_2$Se$_3$ with
short-range interactions. Thanks to the peculiar Dirac band structure, we
find a specific odd-parity triplet pairing is favored in a wide parameter
range, giving rise to a topological superconductor. We propose an unusual
flux quantization in Josephson interferometry as a sharp test of such a
pairing symmetry. We also explicitly demonstrate the existence of gapless
surface Andreev bound states in the resulting topological phase.

We start by introducing Nambu notation $\xi^\dagger_{\mathbf{k}} \equiv [ c_{%
\mathbf{k}, a\alpha} ^\dagger, c_{{-\mathbf{k}}, a \beta}
^T (i s_y )_{\beta \alpha} ]$, where $\alpha, \beta=\uparrow, \downarrow$
label electron's spin and $a$ labels the orbital basis for cell-periodic
Bloch wave-functions. The BCS mean-field Hamiltonian $H=\int_{BZ} d \mathbf{k%
} \; \xi^\dagger_{\mathbf{k}} \mathcal{H}(\mathbf{k}) \xi_{\mathbf{k}} $
uniquely defines a BdG Hamiltonian
\begin{equation}
\mathcal{H}(\mathbf{k})= [ H_0(\mathbf{k}) - \mu ] \tau_z + \hat{\Delta}(%
\mathbf{k})\tau_x,
\end{equation}
where $H_0$ describes the band structure of normal metal, $\mu$ is chemical
potential, and $\hat{\Delta}$ is pairing potential. For TRI superconductors,
$\Theta \mathcal{H}(\mathbf{k})\Theta^{-1} = \mathcal{H}(-\mathbf{k})$ where
$\Theta = is_y K$ is time reversal operation.

The BdG Hamiltonian $\mathcal{H(\mathbf{k})}$ of a fully gapped
superconductor, which describes Bogoliubov quasi-particle spectrum, formally
resembles the Bloch Hamiltonian of an insulator. An important difference,
however, is that $\mathcal{H(\mathbf{k})}$ has particle-hole symmetry
inherited from the doubling of degrees of freedom in Nambu space: $\Xi
\mathcal{H}(\mathbf{k}) \Xi = - \mathcal{H}(-\mathbf{k})$ with $\Xi \equiv
s_y \tau_y K$. Because of this extra symmetry, Schnyder, Ryu, Furusaki and
Ludwig\cite{ucsb} and Kitaev\cite{kitaev} have shown that 3D TRI
superconductors are mathematically classified by an integer invariant $n$
instead of $Z_2$ invariants for insulators\cite{fkm, balents}. Despite this
difference, since $\mathcal{H(\mathbf{k})}$ belongs to a subset of TRI
Hamiltonians, we observe that $\nu \equiv n$ mod $2$ is nothing but its own $%
Z_2$ invariant as explicitly defined in Ref.\cite{fk}. It then follows that $%
\nu=1$ implies a nonzero $n$ and is sufficient (though not necessary) to
establish a topological superconductor phase.

A powerful ``parity criterion'' has been advanced by Fu and Kane to evaluate
$\nu$ efficiently for materials with inversion symmetry\cite{fukane}. This
motivates us to study topological superconductors in centrosymmetric
materials, for which the pairing symmetry can be either even or odd under
inversion. It follows from the explicit formula for $n$\cite{ucsb} that
even-parity ones cannot be topological superconductors. In this work we
focus on odd-parity superconductors satisfying $P H_0(\mathbf{k}) P = H_0(-%
\mathbf{k})$ and $P \hat{\Delta}(\mathbf{k}) P = - \hat{\Delta}(-\mathbf{k})
$, where $P$ is inversion operator. We now provide a simple criterion for
odd-parity topological superconductors:

\textbf{Criterion:} a fully gapped TRI superconductor with odd-parity
pairing is a topological superconductor, if its Fermi surface encloses an
odd number of TRI momenta in the Brillouin zone.

A special case of this criterion has been proved\cite{sato} for certain
triplet superconductors in which $H_{0}(\mathbf{k})=H_{0}(-\mathbf{k})$ and $%
\hat{\Delta} (\mathbf{k})=- \hat{\Delta} (-\mathbf{k})$, i.e., inversion
simplifies to an identity operator $P=I 
$. Here we generalize the proof to all odd-parity superconductors, as needed
later. \textit{Proof}: Since $P \mathcal{H}(\mathbf{k}) P \neq \mathcal{H}(-%
\mathbf{k})$, the parity criterion of Ref.\cite{fukane} does not apply
directly. Instead, because $\hat{\Delta} \tau _{x}$ anticommutes with $\tau
_{z}$, $\mathcal{H}(\mathbf{k)}$ satisfy:
\begin{equation}
\tilde{P}\mathcal{H}(\mathbf{k})\tilde{P}=\mathcal{H}(-\mathbf{k}),\;\;%
\tilde{P}\equiv P\tau _{z}.  \label{inversion}
\end{equation}
Since the operator $\tilde{P}$ defined here satisfies $\tilde{P} ^{2}=1$ and
$[\tilde{P},\Theta ]=0$, $\tilde{P}$ can be used in place of $P$ as an
inversion operator for odd-parity superconductors. The corresponding parity
criterion with $\tilde{P}$ reads
\begin{equation}
(-1)^{\nu }=\prod_{\alpha,m}\xi _{2m}(\Gamma _{\alpha }).  \label{nu}
\end{equation}
Here $\Gamma _{\alpha}$'s ($\alpha =1,...8$) are eight TRI momenta in 3D
Brillouin zone satisfying $\Gamma_{\alpha}= - \Gamma_{\alpha}$ up to a
reciprocal lattice vector. $\xi _{2m}(\Gamma _{\alpha })=\pm 1$ is the $%
\tilde{P}$ eigenvalue of the $2m $-th negative energy band at $\Gamma
_{\alpha }$, which shares the same value $\xi _{2m}(\Gamma _{\alpha })=\xi
_{2m+1}(\Gamma _{\alpha })$ as its Kramers degenerate partner. The product
over $m$ in (\ref{nu}) includes all negative energy bands of $\mathcal{H}(%
\mathbf{k})$. The physical meaning of (\ref{nu}) becomes transparent in
weak-coupling superconductors, for which the pairing potential is a small
perturbation to $H_{0}$. As long as the bands $\varepsilon_{n}(\Gamma
_{\alpha })$ of $H_{0}(\Gamma _{\alpha })$ stay away from the Fermi energy
(which is generically true), pairing-induced mixing between electrons and
holes in the eigenstates $\psi _{m}(\Gamma _{\alpha })$ of $\mathcal{H}%
(\Gamma _{\alpha })$ can be safely neglected. So we have $\xi _{2m}(\Gamma
_{\alpha})=p_{2m}(\Gamma _{\alpha })\times \tau _{2m}(\Gamma _{\alpha })$
with $p$ and $\tau $ being the eigenvalues of $P$ and $\tau _{z}$
separately. (\ref{nu}) then factorizes into two products over $p$ and $\tau $%
. Now the key observation is that the set of all negative energy eigenstates
of $\mathcal{H }$ corresponds to the set of \textit{all} energy bands of $%
H_{0}$ (both above and below $\mu$), which form a complete basis of $H_{0}$.
So we find $\prod_{m}p_{2m}(\Gamma _{\alpha })=\mathrm{Det}[P]=\pm 1$
independent of $\Gamma _{\alpha }$, and thus $\prod_{\alpha,m}p_{2m}(\Gamma
_{\alpha })=1$. (\ref{nu}) then simplifies to
\begin{equation}
(-1)^{\nu }=\prod_{\alpha ,m}\tau _{2m}(\Gamma _{\alpha })=\prod_{\alpha
}(-1)^{N(\Gamma _{\alpha })}.  \label{n}
\end{equation}
Here $N(\Gamma _{\alpha })$ is defined as the number of unoccupied bands at $%
\Gamma _{\alpha }$ in the normal state. (\ref{n}) now has a simple
geometrical interpretation: $\nu =0$ or $1$ if the Fermi surface of $H_{0}$
encloses an even or odd number of TRI momenta, respectively\footnote{%
Strictly speaking, this relation between $\nu$ and Fermi surface topology
only holds for weak-coupling superconductors.}. The latter case corresponds
to a topological superconductor.

A well-known example of odd-parity pairing is superfluid He-$3$\cite{leggett}%
. In particular, the TRI and fully-gapped $B$-phase has been recently
identified as a topological superfluid \cite{ucsb, stanford, roy}, in
agreement with the above criterion. This identification explains the
topological origin of its gapless surface Andreev bound states theoretically
predicted before\cite{he3}. Odd-parity pairing in superconductors is less
well established. A famous example is Sr$_{2}$RuO$_{4}$, as shown by
phase-sensitive tests of pairing symmetry\cite{sr2ruo4}. However, the
observed signatures of spontaneous time reversal symmetry breaking\cite{kerr}
seem to prevent Sr$_{2}$RuO$_{4}$ from being a TRI topological
superconductor.

In the search for odd-parity superconductors, we turn our attention to the
newly discovered superconductor Cu$_{x}$Bi$_{2}$Se$_{3}$---a doped
semiconductor with low electron density and $T_{c}=3.8K$\cite{cubise}.
A most recent angle-resolved photoemission spectroscopy experiment\cite{arpes} found
that the dispersion $\varepsilon_{\mathbf{k}}$ near center
of the Brillouin zone $\Gamma$ strikingly resembles a massive 3D
Dirac fermion, being quadratic near the band bottom,
and linear at higher energy. Upon doping with Cu, the Fermi energy
moves into the conduction band, about $0.25$eV above
the band bottom in the \textquotedblleft relativistic\textquotedblright\ linear regime\cite{arpes}. 
To the best of our knowledge, this is the first discovery of superconductivity in a
3D Dirac material, which motivates us to study its pairing
symmetry.

The Dirac band structure in the parent compound Bi$_{2}$Se$_{3}$ originates
from strong inter-band SOC and can be understood from $k\cdot p$ theory\cite%
{bitetheory}. Since the (lowest) conduction and (highest) valence band at $%
\Gamma $ have opposite parity, general symmetry considerations show that the
$k\cdot p$ Hamiltonian $H_{0}(\mathbf{k})$ to first order in $k$ takes the
form of a four-component Dirac Hamiltonian\cite{dirac}:
\begin{equation}
H_{0}(\mathbf{k})=m\Gamma _{0}+v(k_{x}\Gamma _{1}+k_{y}\Gamma
_{2})+v_{z}k_{z}\Gamma _{3},  \label{H0}
\end{equation}%
where $\Gamma_i$'s ($i=0,\dots,3$) are $4\times 4$ Dirac Gamma matrices.
The four components arise from electron's orbital ($\sigma$) and spin ($s$).
As shown by first-principle calculations\cite{bisbbise, bitetheory},
the conduction and valence bands of Bi$_{2}$Se$_{3}$
mainly consist of two orbitals: the top and bottom Se $p_{z}$-orbital in the five-layer unit
cell, each mixed with its neighboring Bi $p_{z}$-orbital ($z$ is along $c$ axis).
The two orbitals transform into each other under inversion and we label them by $\sigma_z=\pm 1$.
The Gamma matrices in $H_0$ are then expressed as follows: $\{ \Gamma_0, \Gamma_1, \Gamma_2, \Gamma_3
\} \equiv \{ \sigma_x, \sigma_z \otimes s_y, -\sigma_z \otimes s_x, \sigma_y
\}$.

To study superconductivity in Cu$_x$Bi$_2$Se$_3$, 
we consider the following phenomenological effective Hamiltonian with
short-range density-density interactions:
\begin{equation}
H_{\mathrm{eff}} = c^\dagger (H_0 - \mu) c - \int dx \left[ U \sum_{a=1,2}
n_a^2 + 2V n_1n_2 \right],  \label{uv}
\end{equation}
where $n_a(x)=\sum_{\alpha=\uparrow, \downarrow}c_{a\alpha}^\dagger(x)
c_{a\alpha}(x)$ is electron density in orbital $a$. $U$ and $V$ are
intra-orbital and inter-orbital interactions, respectively. All other local
interaction terms, such as $(c^\dagger \sigma_x c)^2$ and $(c^\dagger
\sigma_x\vec{s} c)^2$, are neglected\cite{comment-neglect}. $H_{\mathrm{eff}%
} $ is to be thought of as an effective low-energy Hamiltonian, which
includes the effects of both Coulomb and electron-phonon interactions. We
will assume that at least one of them is positive, giving rise to pairing.
Since $U$ and $V$ are difficult to estimate from a microscopic theory, we
will treat them as phenomenological parameters. Naively, one would expect
that the intra-orbital effective phonon-mediated attraction would be
stronger than the inter-orbital one. However, since the same is true for the
Coulomb repulsion, it is possible that the overall effective interactions
satisfy, e.g., $0<V<U$.

\begin{table}[h]
\begin{tabular}{|c|c|c|c|c|}
\hline
$\hat{\Delta}$: & $\Delta_{1} I + \Delta^{\prime}_1 \Gamma_0$ & $\Delta_{2}
\Gamma_{50}$ & $\Delta_{3} \Gamma_{30}$ & $\Delta_4 ( \Gamma_{10} ,
\Gamma_{20} )$ \\ \hline
$\Theta$ & $+$ & $+$ & $+$ & $(+, +)$ \\
$P$ & $+$ & $-$ & $-$ & $(-, -)$ \\
$C_3$ & $z$ & $z$ & $z$ & $(x, y)$ \\
$M$ & $+$ & $-$ & $+$ & $(-, +)$ \\ \hline
\end{tabular}%
\caption{Pairing potential in mean-field BdG Hamiltonian of $U-V$ model, and
their transformation rules. }
\label{tab:SC}
\end{table}

To determine the pairing symmetry of the $U-V$ model, we take advantage of
two facts: a) near $T_{c}$, $\hat{\Delta}$ 
forms an irreducible representation of crystal point group; b) the
mean-field pairing potential is local in $x$ and thus $k$-independent. The
form of all such pairing potentials $\hat{\Delta}$ are listed in Table II,
where $\Gamma _{5}\equiv \Gamma _{0}\Gamma _{1}\Gamma _{2}\Gamma
_{3}=\sigma_z s_z$ and $\Gamma _{jk}\equiv i\Gamma _{j}\Gamma _{k}$. Also
shown are transformation rules of $\hat{\Delta}$'s under the following
symmetry operations of Bi$_{2}$Se$_{3}$: inversion $P=-\Gamma _{0}=-\sigma
_{x}$, threefold rotation around the $c$ axis $C_{3}=\exp (i \Gamma_{12} \pi
/3)=\exp (-is_{z}\pi /3)$, and mirror about $yz$ plane $M=-i
\Gamma_{15}=-is_{x}$. We find that $\hat{\Delta}_{1},...,\hat{\Delta}_{4}$
correspond to $A_{1g},A_{1u},A_{2u}$ and $E_{u}$ representations of point
group $D_{3d}$ respectively. The three $A$ representations are
one-dimensional so that the corresponding phases are non-degenerate. Among
them, $c^{T}\hat{\Delta}_{1}(is_{y})c\propto \Delta _{1}c_{1\uparrow
}c_{1\downarrow }+\Delta _{1}^{\prime }c_{1\uparrow }c_{2\downarrow
}+(1\leftrightarrow 2)$ is spin-singlet pairing with mixed intra- and
inter-orbital (orbital triplet) components, which is invariant under all crystal
symmetries; $c^{T}\hat{\Delta}_{2}(is_{y})c\propto (c_{1\uparrow
}c_{2\downarrow }+c_{1\downarrow }c_{2\uparrow })$ is inter-orbital (orbital
singlet) spin-triplet pairing; $c^{T}\hat{\Delta}_{3}(is_{y})c\propto
(c_{1\uparrow }c_{1\downarrow }-c_{2\uparrow }c_{2\downarrow })$ is
intra-orbital spin-singlet pairing. The $E_{u}$ representation is
two-dimensional with $c^{T}\hat{\Delta}_{4}(is_{y})c\propto \alpha
c_{1\uparrow }c_{2\uparrow }+\beta c_{1\downarrow }c_{2\downarrow }$, where $%
\alpha $ and $\beta $ are arbitrary coefficients, leading to a $SU(2)$
degenerate manifold at $T_{c}$. Of these phases, the $\hat{\Delta}_{2}$
pairing phase is odd-parity, TRI, \textit{and} fully gapped, with a
Bugoliubov spectrum given by%
\begin{equation}
E_{\pm ,k}=\sqrt{\varepsilon _{k}^{2}+\mu ^{2}+\Delta_2 ^{2}\pm 2\mu \sqrt{%
\varepsilon _{k}^{2}+\left( \frac{m}{\mu }\right) ^{2}\Delta_2 ^{2}}}\text{,}
\end{equation}%
where $\varepsilon _{k}=\sqrt{m^{2}+v^{2}\left( k_{x}^{2}+k_{y}^{2}\right)
+v_{z}^{2}k_{z}^{2}}$. Since the Fermi surface only encloses the $\Gamma $
point, according to our earlier criterion $\hat{\Delta}_{2}$ pairing gives
rise to a topological superconductor phase in the $U-V$ model for Cu$_{x}$Bi$%
_{2}$Se$_{3}$.

\begin{figure}[tbp]
\centering
\includegraphics[width=0.4\textwidth]{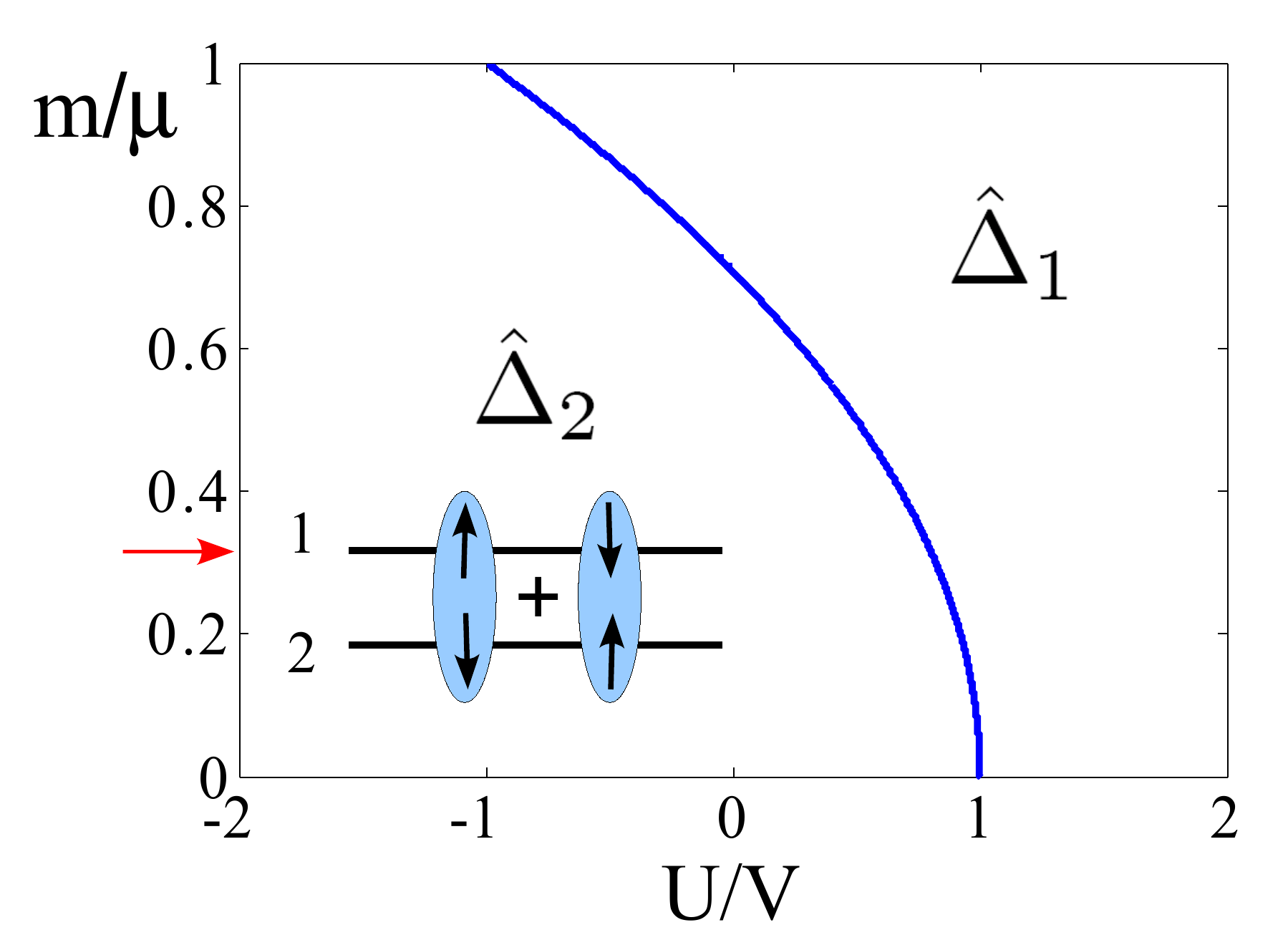}
\caption{Phase diagram of the $U-V$ model, showing the highest $T_c$ phase
as a function of $m/\protect\mu$ and $U/V$ (assuming $V>0$). The arrow shows
the experimental estimate for $m/\protect\mu$, which is about $\frac{1}{3}$
\protect\cite{arpes}.
Two phases $\hat{\Delta}_1$ and $\hat{\Delta}_2$
appear, which are even and odd under parity respectively (see Table \protect
\ref{tab:SC}). The insets shows schematically the structure of the Cooper
pair wavefunction in the $\hat{\Delta}_2$ phase, consisting of two electrons
localized on the top (1) and bottom (2) of the five-atom unit cell.}
\label{fig1}
\end{figure}

We now solve the linearized gap equation for $T_{c}$ of the various pairing
channels to obtain the phase diagram.
For purely inter-obital pairing $\hat{\Delta}_{2}$ and $\hat{\Delta}_{4}$,
the gap equation reads $V\chi _{2,4}(T_{c})=1$. For purely intra-orbital
pairing $\hat{\Delta}_{3}$, it reads $U\chi _{3}(T_{c})=1$. Here $\chi
_{i}(T)$ is the finite temperature superconducting susceptibility in pairing
channel $\hat{\Delta}_{i}$. A straight-forward calculation shows that
\begin{equation}
\chi _{2}=\chi _{0}\int d\mathbf{k}\delta (\varepsilon _{\mathbf{k}}-\mu )%
\mathrm{Tr}[\Gamma _{50}\mathrm{P}_{\mathbf{k}}]^{2}/(2D_{0}).  \label{sc}
\end{equation}%
Here $\chi _{0}\equiv D_{0}\int_{0}^{W}d\varepsilon \tanh \left( \frac{%
\varepsilon }{2T}\right) /\varepsilon $, where $D_{0}$ is density of states
at Fermi energy and $W$ is high-energy cutoff. The projection operator $%
\mathrm{P}_{\mathbf{k}}\equiv \sum_{\lambda =1,2}|\phi _{\lambda ,\mathbf{k}%
}\rangle \langle \phi _{\lambda ,\mathbf{k}}|$ is
defined by the two degenerate Bloch states at $\mathbf{k}$. As we will see,
the integral over the Fermi surface in (\ref{sc}), which takes into account
the interplay between pairing potential and band structure effects in a
\textit{multi-orbital} system, will play a key role in favoring $\hat{\Delta}%
_{2}$ pairing. The other two susceptibilities $\chi _{3}$ and $\chi _{4}$
can be obtained simply by replacing $\Gamma _{50}$ in (\ref{sc}) with $%
\Gamma _{30}$ and $\Gamma _{10}$ respectively. Using $\mathrm{P}_{\mathbf{k}%
}=\frac{1}{2}(1+\sum_{\nu =0}^{3}n_{\mathbf{k}}^{\nu }\Gamma _{\nu })$ and $%
n_{\mathbf{k}}=(m,vk_{x},vk_{y},v_{z}k_{z})/\varepsilon _{\mathbf{k}}$ for
Dirac Hamiltonian $H_{0}$, we obtain $\chi _{2}=\chi _{0}(1-m^{2}/\mu ^{2})$%
, $\chi _{3}=\chi _{4}=2\chi _{2}/3$. 
The gap equation for the intra- and inter-orbital mixed pairing $\hat{\Delta}%
_{1}$ is:
\begin{equation}
\det \left[ \left(
\begin{array}{cc}
U\chi _{0} & U\chi _{0}C_{1} \\
V\chi _{0}C_{1} & V\chi _{0}C_{2}%
\end{array}%
\right) -I\right] =0,  \label{gp}
\end{equation}%
%
where $C_{n}=(m/\mu )^{n}$ for $n=1,2$. From (\ref{sc}) and (\ref{gp}), we
now deduce the phase diagram. 
Since $\chi _{3}<\chi _{0}$ and $\chi _{4}<\chi _{2}$, $\hat{\Delta}_{3}$
and $\hat{\Delta}_{4}$ always have a lower $T_{c}$ than their counterparts $%
\hat{\Delta}_{1}$ and $\hat{\Delta}_{2}$, respectively. Only the latter two
phases appear in the phase diagram. By equating their $T_{c}$'s, we obtain
the phase boundary:
\begin{equation}
U/V=1-2m^{2}/\mu ^{2}.
\end{equation}%
Fig.1 shows the highest $T_{c}$ phase as a function of $U/V$ and $m/\mu $,
for positive (attractive) $V$. The $\hat{\Delta}_{2}$ pairing phase
dominates in a significant part of the phase diagram. Note that
experimentally, it has been estimated that $m/\mu \approx \frac{1}{3}$\cite{arpes}.
When $V<0$ the $\hat{\Delta}_{1}$ phase is stable for $U>m^{2}/\mu
^{2}|V|$, whereas for smaller $U$ the system is non-superconducting. The
fact that the phase boundary starts at the point $U=V$ and $m=0$ is not
accidental: at this point the Hamiltonian (\ref{uv}) has an enlarged $U(1)$
chiral symmetry: $c\rightarrow \exp (i\theta \Gamma _{50})c$. Under
the unitary transformation $ \exp (i\pi \Gamma _{50}/4)$, the two pairing potentials
$c^T  (i s_y) c $ and $c^T \Gamma_{50} (is_y)c$ transform into each other\footnote
{For $m=0$ and arbitrary $U/V$, a $Z_2$ symmetry $c \rightarrow \Gamma_{50} c$
forbids mixing between intra- and inter-orbital components in $\hat{\Delta}_1$ pairing.}.

\begin{figure}[tbp]
\centering
\includegraphics[width=0.37\textwidth]{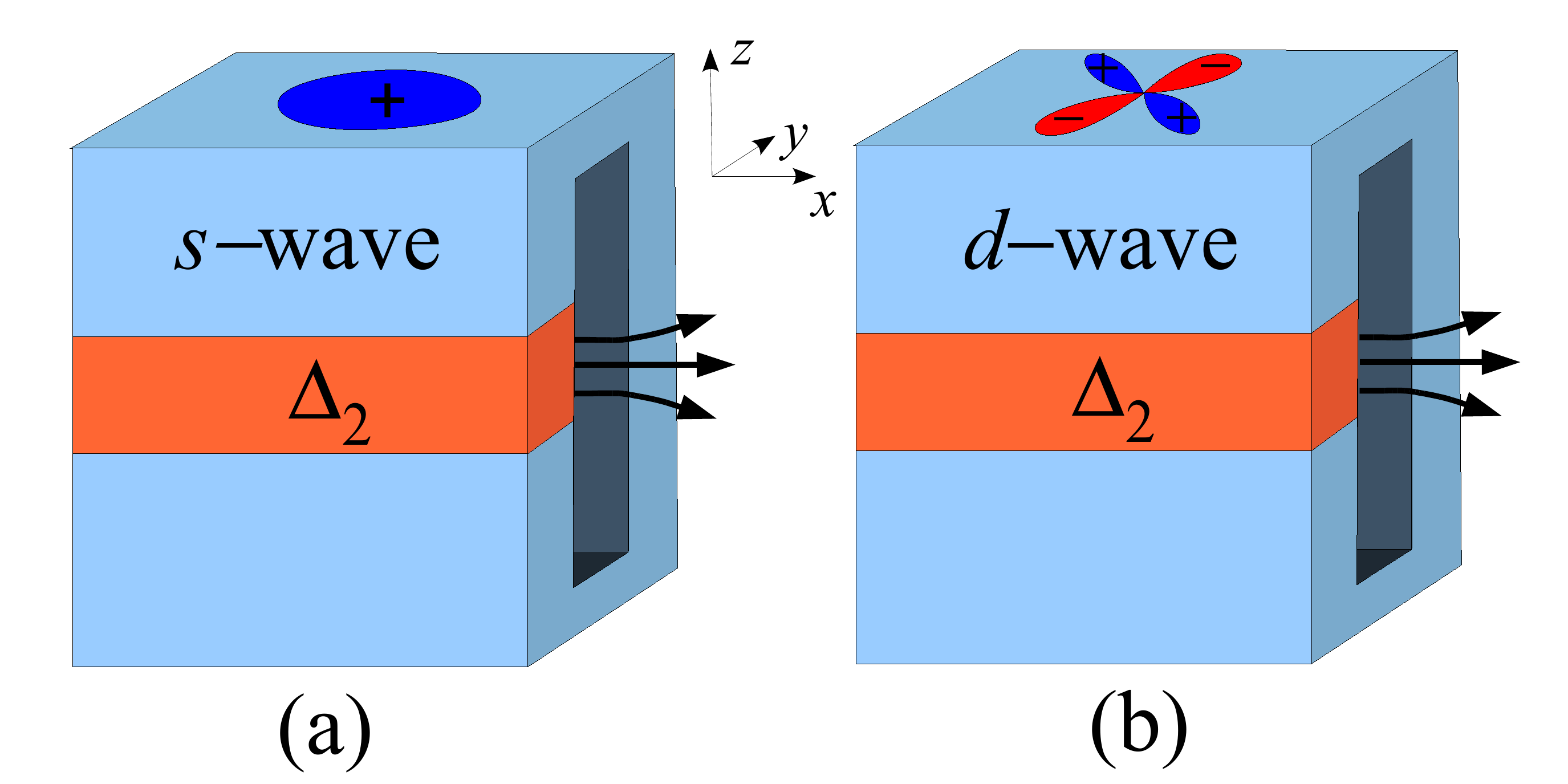}
\caption{ Phase sensitive experiments to test $\Delta_2$ pairing, which is
odd under both inversion ($\mathbf{r} \rightarrow -\mathbf{r}$) and
reflection about the $yz$ plane ($x \rightarrow -x$). A superconducting ring
made of either an $s-$wave (a) or $d-$wave (b) superconductor contains a
segment of a $\Delta_2$ superconductor. The flux through the ring is $nh/4e$
(a) or $(n+\frac{1}{2})h/2e$ (b), where $n$ is an integer.}
\label{fig2}
\end{figure}

From now on, we focus on the topologically non-trivial $\hat{\Delta}_2$
phase. To obtain the surface Andreev bound state spectrum, we solve the BdG
Hamiltonian in a semi-infinite geometry $z<0$:
\begin{eqnarray}
\mathcal{H} (k_x, k_y) &=& (-i v_z \Gamma_3 \partial_z + m \Gamma_0 -\mu)
\tau_z + \Delta_2 \Gamma_{50} \tau_x  \nonumber \\
&+& v ( k_x \Gamma_1 + k_y \Gamma_2) .  \label{H}
\end{eqnarray}
The continuum Hamiltonian (\ref{H}) must be supplemented with a boundary
condition at $z=0$. 
For a Cu$_x$Bi$_2$Se$_3$ crystal naturally cleaved in between two five-layer
unit cells, the wavefunction amplitude on the bottom layer corresponding to $%
\sigma_z=-1$ must vanish, so that $\sigma_z \psi = \psi|_{z=0} \label{bc} $
is satisfied. By solving $\mathcal{H}$ at $k_x=k_y=0$, we find that a
Kramers pair of zero-energy surface Andreev bound states $\psi_{\pm}$ exist
for $\mu^2>m^2 - \Delta_2^2$, i.e., as long as the bulk gap remains finite.
The wavefunctions of $\psi_{\pm}$ are particularly simple for $m=0$\cite%
{zeromode}:
\begin{eqnarray}
\psi_{\pm} (z)&=& e^{-\kappa z}(\cos k_0 z |\sigma_z = 1\rangle + \sin k_0 z
| \sigma_z = -1 \rangle)  \nonumber \\
& & \otimes | s_z = \pm 1, \tau_y =\mp 1 \rangle,
\end{eqnarray}
where $\kappa=\Delta_2/v_z$ 
and $k_0=\mu/v_z$.
Using $k \cdot p$ theory, we obtain the low-energy Hamiltonian describing
the dispersion of surface Andreev bound states at small $k_x$ and $k_y$: $%
H_{sf}= v_s ( k_x s_y - k_y s_x). $ The velocity $v_s$ is given by $%
v_s=\langle \psi_- | v \Gamma_1 \tau_z | \psi_+\rangle/\langle \psi_+ |
\psi_+ \rangle \simeq v \Delta_2^2 / \mu^2$.

%

Finally, we discuss the experimental consequences of the $\Delta_2$ state.
The topologically protected surface state 
can be detected by scanning tunneling microscopy. In addition, the oddness
of this state under parity and mirror symmetries has consequences for
phase-sensitive experiments. Consider a c-axis Josephson junction
between a $\Delta_2$ superconductor and an $s-$wave superconductor. Since
the $\Delta_2$ state is odd under reflection about the $yz$ plane, whereas
the $s-$wave gap $\Delta_s$ is even, the leading order Josephson coupling
between the two superconductors, $-J_1(\Delta_s^* \Delta_2+c.c.)$, vanishes
(as well as higher odd-order terms).
The second order Josephson coupling, $-J_2[(\Delta_s^*)^2 \Delta_2^2+c.c.]$,
can be non-zero (as well as higher even-order terms). Therefore, the flux
through a superconducting ring shown in Fig. 2a is quantized in units of $%
\frac{h}{4e}$\cite{Berg}. Alternatively, in a Josephson junction between a $%
\Delta_2$ superconductor and a $d-$wave superconductor oriented as shown in
Fig. 2b, 
the first order Josephson coupling is non-zero. The flux through the ring in
Fig. 2b takes the value $\frac{h}{2e}(n+\frac{1}{2})$ ($n$ is an integer).
The same holds for an $s-$wave superconductor which does not have the mirror
symmetry relative to the $yz$ plane. The observation of these anomalous flux
quantization relations would be a unique signature of the topological $%
\Delta_2$ state.

To conclude, we present a theory of odd-parity topological superconductors
and propose the newly discovered superconductor Cu$_x$Bi$_2$Se$_3$ as a
potential candidate for this new phase of matter. We hope this work will
bridge the study of topological phases and unconventional superconductivity,
as well as stimulate the search for both in centrosymmetric materials with
spin-orbit coupling.

It is a pleasure to thank Hsin Lin, Charlie Kane, Steve Kivelson and
especially Patrick Lee for very helpful discussions. We are grateful to
Zahid Hasan and Andrew Wray for sharing their data prior to publication.
This work was supported by the Harvard Society of Fellows (LF) and by the
National Science Foundation under grants DMR-0705472 and DMR-0757145 (EB).


\begin{thebibliography}{99}
\bibitem{comment-neglect} Microscopically, these additional terms are
proportional to the square of the wavefunction overlap of orbitals localized
on the top and bottom of the five-atom unit cell, about $9\mathrm{\mathring{A%
}}$ apart. Therefore they are expected to be small.
\end{thebibliography}
\end{document}